\documentclass[12pt]{article}
\usepackage{geometry}
\usepackage{a4}
\usepackage{graphicx}
\usepackage{epsf}
\usepackage{amsmath}
\usepackage{amssymb}
\usepackage{cite}
\newcommand{\be}{\begin{equation}}
\newcommand{\ee}{\end{equation}}

\newcommand{\Rmnum}[1]{\expandafter\@slowromancap\romannumeral #1@}
\newcommand{\bea}{\begin{eqnarray}}
\newcommand{\eea}{\end{eqnarray}}
\begin{document}
\def\C{{\mathbb{C}}}
\def\R{{\mathbb{R}}}
\def\s{{\mathbb{S}}}
\def\T{{\mathbb{T}}}
\def\Z{{\mathbb{Z}}}
\def\W{{\mathbb{W}}}
\def\Bbb{\mathbb}
\def\BZ{\Bbb Z} \def\BR{\Bbb R}
\def\BW{\Bbb W}
\def\BM{\Bbb M}
\def\BC{\Bbb C} \def\BP{\Bbb P}
\def\CP{\BC\BP}
\begin{titlepage}
\title{Scalar Radiation in the Background of a Naked Singularity} \author{} 
\date{%
Anshuman Dey, Pratim Roy, Tapobrata Sarkar 
\thanks{\noindent 
E-mail:~ deyanshu, proy, tapo@iitk.ac.in} 
\vskip0.4cm 
{\sl Department of Physics, \\ 
Indian Institute of Technology,\\ 
Kanpur 208016, \\ 
India}} 
\maketitle 
\abstract{We study scalar radiation spectra from a particle in circular orbit, in the background of the Janis-Newman-Winicour (JNW) naked singularity. 
The differences in the nature of the spectra, from what one obtains with a Schwarzschild black hole, is established. 
We also compute the angular distribution of the spectra.
}
\end{titlepage}

\section{Introduction}

Naked singularities \cite{waldbook} have been objects of great interest in the theory of general relativity for some time now. With a proof of the
cosmic censorship hypothesis still not in place, attempts have been made to understand possible realistic physical scenarios involving these objects. 
For example, the work of \cite{ellis} models galactic centers with naked singularities, and studies them in the context of gravitational 
lensing (see also \cite{Perlick}). On the other hand, efforts have been made to confirm that these are stable objects, and there are hints from numerical 
work that this might be the case \cite{shapteu} (more recent attempts to settle the question of stability of naked singularites were made in \cite{theo}). Recently, the
Janis-Newman-Winicour (JNW) naked singularity \cite{JNWrefs},\cite{virbhadra} has also been shown to be stable under 
scalar field perturbations\cite{Varadarajan}. 

It is important to gain knowledge about naked singularities, and to 
contrast them with their black hole cousins. Several works in this direction appear in the literature, but to the best of our knowledge, there has so far been no 
attempt to calculate radiation spectra in naked singularity backgrounds. 
In this paper, we study the scalar power radiation spectrum in the background of the JNW naked singularity. Power spectra in black hole backgrounds 
have been studied in several works (see, e.g \cite{Misner}, \cite{Kerr}, \cite{book}) using perturbation theory. 
Here, following \cite{Misner} (see \cite{cardoso} for a more recent treatment in the context of AdS black holes), we consider a scalar field 
perturbation \cite{Varadarajan} of the JNW naked singularity, with a massive particle coupled to this field,
and calculate the scalar power radiated. Mostly in the literature, such work has been done in connection with
synchrotron radiation, i.e  where most of the radiated power lies in a range of frequencies that is large compared to the orbital frequency of the particle.
In fact, the work of \cite{Misner} was initiated with the hope that such 
radiation would be observable, although later work \cite{davis} suggested limitations on the astrophysical implications of gravitational synchrotron 
radiation.\footnote{In a slightly different context, radiation from particles in orbit around black holes has been investigated in a series of papers \cite{poisson}.} 
Here, we present a comprehensive analysis of the scalar power spectrum emitted by a particle in a circular orbit, coupled to a scalar field, 
in the JNW background. We do this both for stable as well as unstable orbits, by employing a parabolic WKB approximation. We find that for a certain range 
of parameters, the radiation spectrum is different from what is obtained with black hole backgrounds. For a different range of parameters, there are 
two distinct types of spectra, one of which mimics a black hole. 

This paper is organized as follows. In section 2, we review the physics of the JNW metric and set up the basic equations 
for calculating the scalar power. Section 3 discusses the methodology and our main results, and we conclude in section 4 with some discussions. 

\section{The JNW Metric and the Scalar Equations}

In this section, we start by reviewing the basic features of the JNW metric and circular geodesics therein, following the work of \cite{patil1}. 
We then proceed to set up the equations for scalar radiation in the background of this metric, which will enable us to calculate the 
power spectrum in the next section. Using the notation of \cite{virbhadra}, we start with the JNW metric \cite{JNWrefs} :
\begin{equation}
ds^2 = -\left(1-\frac{b}{r}\right)^\nu dt^2 + \frac{1}{\left(1-\frac{b}{r}\right)^\nu}dr^2 + r^2\left(1-\frac{b}{r}\right)^{1-\nu}\left(d\theta^2+\sin^2\theta d\phi^2\right)
\label{a1}
\end{equation}
Here, $\nu$ varies between $0$ and $1$. It can be verified that this space-time has no horizon. Further, it can be shown
\cite{joshi} that it satisfies the weak energy conditions and that the singularity is globally naked. 
The JNW space-time is sourced by a scalar field,
\begin{equation}
 \psi = \frac{q}{b\sqrt{4\pi}}\ln\left(1-\frac{b}{r}\right)
 \label{a2}
\end{equation}
where $q$ is its magnitude. The ADM mass of the space-time $M$ is related to the parameters $b$ and $q$ by 
$b = 2\sqrt{q^2 + M^2}$, and the parameter $\nu$ appearing in eq.(\ref{a1}) is given by $\nu = 2M/b$. The Schwarzschild limit of the JNW singularity
is obtained by taking $\nu = 1$, i.e $q=0$.
We will specialize to the case where we have a particle moving in a circular orbit in the JNW geometry, setting $\theta = \pi/2$, using the
spherical symmetry of the system. The geodesic equations appear in \cite{patil1} and we briefly
review them. For time-like geodesics, the equations following from the metric are,
\begin{equation}
{\dot t} =  \gamma \left(1-\frac{b}{r}\right)^{-\nu}
 \label{a6}
\end{equation}
\begin{equation}
\dot{\phi}=  \frac{h}{\left(1-\frac{b}{r}\right)^{1-\nu} r^2}
 \label{a5}
\end{equation}
\begin{equation}
 \dot{r}^2+W(r) = 0
 \label{a3}
\end{equation}
where,
\begin{equation}
 W(r) = \left(1-\frac{b}{r}\right)^\nu \left[1+\frac{h^2}{r^2(1-\frac{b}{r})^{1-\nu}}\right] - \gamma^2
 \label{a4}
\end{equation}
The dots denote differentiation with respect to an affine parameter, and $h$ and $\gamma$ are conserved quantities, corresponding
to the angular momentum and the total energy. The angular frequency of the particle is
\begin{eqnarray}
\omega_0 = \frac{d\phi}{dt} = \frac{h}{r^2 \gamma} \left(1-\frac{b}{r}\right)^{2\nu-1}
\label{a7}
\end{eqnarray}
For a circular geodesic, $\dot{r} = 0$, and hence from eqs. (\ref{a3}) and (\ref{a4}),
\begin{eqnarray}
\gamma^2 = W(r_0)= \left(1-\frac{b}{r_0}\right)^\nu \left[1+\frac{h^2}{r_0^2(1-\frac{b}{r_0})^{1-\nu}}\right] 
\label{a8}
\end{eqnarray}
where $r_0$ denotes the radius of the orbit. After imposing $\frac{dW}{dr}|_{r=r_0} = 0$, we obtain
\begin{equation}
\gamma = \left[\frac{\left(1-\frac{b}{r_0}\right)^{\nu} \left(2r_0 - b\nu -b\right)}{\left(2r_0 - 2b\nu - b\right)}\right]^{1 \over 2},~~~
h = r_0\left[\frac{b\nu\left(1 - \frac{b}{r_0}\right)^{1-\nu}}{2r_0 - 2b\nu -b}\right]^{1 \over 2}
\label{hgamma}
\end{equation}
From eq.(\ref{hgamma}), we see that the radius of the photon sphere \cite{ellis}, where the energy and the angular momentum diverges, is at 
\begin{equation}
r_{\rm ps} = \frac{b}{2}(1 + 2\nu)
\label{ps}
\end{equation}
Depending on whether the photon sphere exists (i.e is above $r=b$) or not, the singularity is called a weakly naked or strongly naked 
singularity \cite{ellis}. From eq.(\ref{ps}), we see that for $\nu > 1/2$, the JNW solution becomes a weakly naked singularity, while it remains strongly
naked for $0 < \nu < 1/2$. The quantity $r_{\rm ps}$ will be important for our discussion on the power spectrum. From eq.(\ref{hgamma}), we get
an useful expression for the angular frequency of the particle :
\begin{equation}
\omega_0 = \frac{1}{r_0} \left(1-\frac{b}{r_0}\right)^{(2\nu-1)/2} \left(\frac{b\nu}{2r_0-b\nu-b}\right)^{1/2} 
\label{a11}
\end{equation}
It is also necessary for our purpose to derive some restrictions relating to the radius of the orbit, $r_0$. Such restrictions 
for stable circular orbits have been worked out in \cite{patil1} and we briefly state their results. Using the notation
\begin{equation}
 r_{\pm} = \frac{b}{2}\left(1+3\nu\pm\sqrt{5\nu^2-1}\right)
\label{a17}
\end{equation}
for $0 < \nu < \frac{1}{\sqrt{5}}$, for a stable circular orbit, we require
\begin{equation}
 b < r_0 < \infty
 \label{a18}
\end{equation}
When $\frac{1}{\sqrt{5}} < \nu < \frac{1}{2}$, there are two different regions where stable circular orbits exist:
\begin{equation}
 b < r_0 < r_- \hspace{4mm} \mbox{and} \hspace{4mm}  r_+ < r_0 < \infty  
 \label{a19}
\end{equation}
Finally, for $\frac{1}{2} < \nu < 1$, stable circular gedesics exist for
\begin{equation}
 r_+ < r_0 < \infty
 \label{a20}
\end{equation}
For future reference, we record the expression for $u^0 = \frac{dt}{d\tau}$,
\begin{eqnarray}
  u^0 = \left(1-\frac{b}{r_0}\right)^{-\nu/2} \left(\frac{2r_0 - b\nu - b}{2r_0-2b\nu-b}\right)^{1/2}
 \label{a21}
\end{eqnarray}

Now, in the JNW space-time, we introduce a scalar field $\Phi$,  that does not interact with the field $\psi$ (sourcing the fixed JNW metric), and treat $\Phi$ as
a perturbation over the JNW background. 
That the JNW metric is perturbatively stable under such a scalar field has been demonstrated in \cite{Varadarajan}. 
A massive particle is coupled to the field $\Phi$, and we study scalar emission from this particle. To describe the physics, we write down, following \cite{Misner},
the action describing the interaction of the particle with the field $\Phi$, in the JNW background,
\begin{equation}
S=-\frac{1}{8\pi}\int d^4 x\sqrt{-g}\hspace{1mm} \Phi_{,\mu}\Phi^{,\mu}-m_0\int d\tau(1+q_s\Phi)\sqrt{-\dot{z}^\mu\dot{z}_\mu}
\label{a22}
\end{equation}
Here, $z^{\mu}(\tau)$ denotes the world line coordinates of the particle, $q_s$ its scalar charge and $m_0$
its mass.  The relevant part of the energy momentum tensor is,
\begin{equation}
 T^{\mu \nu} = \frac{1}{4 \pi} \left(\Phi^{, \mu}\Phi^{,\nu}-\frac{1}{2}g^{\mu \nu}\Phi_{, \eta}\Phi^{, \eta}\right)+m_0 (-g)^{1/2} \int{d\tau (1+q_s \Phi)
 \delta^4 (x-z)\dot{z}^\mu \dot{z}^\nu}
 \label{a23}
\end{equation}
The equation of motion for $\Phi$ is 
\begin{equation}
  \frac{\partial}{\partial x^\mu} \left(\sqrt{-g} g^{\mu\nu} \frac{\partial \Phi}{\partial x^\nu}\right)=4 \pi q_s m_0\left(\frac{dz^0}{d\tau}\right)^{-1} \delta^3 (x-z(t))
  \label{a24}
\end{equation}
The delta function in the preceding two equations can be expanded as \cite{Misner} :
\begin{equation}
 \delta^3(x-z(t))=\delta(r-r_0)\delta(\theta-\pi/2)\delta(\phi-\omega_0 t) 
 \label{a25}
\end{equation}
Using the spherical symmetry of the JNW background, we can write
\begin{equation}
 \Phi= \sum_{m=-\infty}^{\infty}   \sum_{l=\vert m\vert}^{\infty} r^{-1} u_{lm}(r)Y^m_l(\theta,\phi)e^{-imw_0t}
 \label{a26}
\end{equation}
and consequently, we obtain the radial equation for $u_{l m}(r)$ as,
\begin{equation}
 -\frac{d^2u_{lm}}{dr_*^2}+\left[\left(1-\frac{b}{r}\right)\left(\frac{b}{r^3}+\frac{l(l+1)}{r^2}\right)-
 m^2\omega_0 ^2\left(1-\frac{b}{r}\right)^{2\left(1-\nu\right)}\right]u_{lm} = C_{lm}\delta(r_*-r_{0*})
 \label{a27}
\end{equation}
where, 
\begin{equation}
  C_{lm}=-4\pi(u^0r_0)^{-1}q_sm_0Y_l^m(\pi/2,0)
  \label{a28}
\end{equation}
and the Regge-Wheeler coordinate $r_{*}$ is given by;
\begin{equation}
 \frac{d r}{d r_{*}} = 1-\frac{b}{r} = f(r)
 \label{a29}
\end{equation}
We will focus on the solution of eq.(\ref{a27}), and use it in conjunction with the power radiated, given by \cite{Misner}
\begin{equation}
 P = \mod{\int{d\Omega r^2 T_t^r}} ~~~= \frac{1}{4 \pi}\int{d\Omega r^2 \frac{\partial \Phi}{\partial t}} \frac{\partial \Phi}{\partial r}
\label{a30}
\end{equation}
Having elaborated on the basic setup, we now proceed to calculate the scalar power spectrum in the JNW background. 

\section{Scalar Power Radiation Spectra in the JNW Background}

First, we find the solution for the homogeneous part of the radial equation of eq.(\ref{a27}). Then, the full solution can be found by a 
Green's function technique. Following \cite{Varadarajan}, the asymptotic behaviour of the homogeneous part of the radial equation is,
\begin{equation}
 u_{lm} \sim A r_* + B ; ~~~r_* \rightarrow -\infty
 \label{a32}
\end{equation}
and
\begin{equation}
 u_{lm} \sim C e^{i m w_0 r_*}+De^{-i m w_0 r_*} ; ~~~r_* \rightarrow \infty
 \label{a33}
\end{equation}
Here, $A$, $B$, $C$ and $D$ are constants. Let $u_L$ and $u_R$ be the left moving and right moving solutions of the radial equation. We
can then write the boundary conditions 
\begin{eqnarray}
 u_L  &\sim&  \begin{cases} (k_\infty)^{-1/2} (e^{-i k r_*}+{\mathcal S} e^{i k r_*});~~~ & r_* \rightarrow + \infty \\ A r_* + B;~~~ & r_* \rightarrow - \infty
\end{cases}
\label{a34}
\\
u_R &\sim& \begin{cases} (k_\infty)^{-1/2} {\mathcal T} e^{i k r_*};~~~ & r_* \rightarrow + \infty \\ C r_*;~~~ & r_* \rightarrow - \infty
\end{cases}
\label{a35}
\end{eqnarray}
where ${\mathcal S}$ and ${\mathcal T}$ are analogues of reflection and transmission coefficients, and $k_{\infty}$ is a normalization factor. The Wronskian is computed as
\begin{equation}
W(+ \infty) = -2 i {\mathcal T}, ~~~~W(- \infty) = -B C
 \label{a36}
\end{equation}
The Green's function is,
\begin{equation}
 G(r_*,r_{0 *})= \begin{cases} \frac{i}{2 {\mathcal T}} u_L(r_{0 *})u_R(r_*);~~~ & r_* > r_{0 *} \\ -\frac{1}{B C} u_R (r_{0 *}) u_L (r_*);~~~ & r_* < r_{0 *}
\end{cases}
\label{a37}
\end{equation}
From eqs.(\ref{a34}) and (\ref{a35}), the main difference of the asymptotic solutions here, as compared to the black hole case can be seen. Namely,
near $r=b$, the solution is not a plane wave. A similar situation occurs in the case of AdS black holes, near spatial infinity \cite{cardoso}, and the wave
function is set to zero there. 
That the situation demands care in the choice of initial conditions as $r \rightarrow b$ is of interest by itself, and deserves further study. 
We will confine ourselves to the calculation of the power radiated towards infinity. Using standard properties of the Green's function of eq.(\ref{a37})
in eqs.(\ref{a22}) and (\ref{a30}), the expression can be shown to be \cite{Misner}
 \begin{equation}
  P = \frac{1}{8\pi} \sum_{m=-\infty}^{\infty}   \sum_{l=\vert m\vert}^{\infty}m\omega_0\vert C_{lm}\vert^2\vert u_L (r_{0*})\vert ^2
  \label{a38}
 \end{equation}
Next, we attempt to solve the radial equation, and hence find an analytic solution for $u_L$ by a WKB approximation. 
Specifically, we use the parabolic WKB method, described in \cite{Misner}. We will very briefly recapitulate the method here, and point out a few
issues regarding its application in our case. 
Let us suppose we have an equation of the form,
\footnote{In the present discussion, we use the variable $x$ to keep the discussion general, following \cite{Misner}. We keep in mind that $x$ 
will be taken as $r_*$ in our case.}
\begin{equation}
\frac{d^2y}{dx^2}+\left[E-V\right]y(x)=0
 \label{a39}
\end{equation}
Further, we assume that the potential can be represented near its maxima $x_0$ by a parabola, $V(x_0 + x) \approx V(x_0) + \frac{1}{2}V^{''}(x_0)x^2 $,
with $V^{''}(x_0) < 0$. Using the parabolic approximation for the potential in eq.(\ref{a39}),
the general expression for the wave function in the region near the maximum of the potential barrier is given by \cite{Misner}
\begin{equation}
u_L\left(x\right)=\left(\frac{1}{2w}\right)^{-\frac{1}{4}}\frac{\Gamma(-\nu)}{\sqrt{2\pi}}\exp\left(-\frac{1}{8}\pi\epsilon-\frac{1}{8}i\pi-
\frac{1}{4}i\epsilon\ln\frac{1}{2}\epsilon\right)D_{\nu}(-\eta x)
 \label{a41}
\end{equation}
where 
\begin{eqnarray}
w = \vert\frac{1}{2}V^{''}(x_0)\vert ^{\frac{1}{2}};~~~\epsilon &=& \frac{V(x_0) -E}{w};~~~\eta = (1-i)w^{\frac{1}{2}};~~~\nu = -\frac{1}{2}(1+i\epsilon)
 \label{a43}
\end{eqnarray}
$D_{\nu}(\eta x)$ is the parabolic cylinder function.
We choose the potential maxima in the parabolic WKB method to be near $x_0 = 0$. In defining the tortoise 
coordinate, we have the freedom to choose the integration constant such that the potential maxima does indeed occur near $r_{0*} =0$. 
Then, in eq.(\ref{a38}), we can use
\begin{eqnarray}
|u_L(0)|^2 \simeq\frac{1}{\sqrt{2w}\pi}e^{-\frac{1}{4}\pi\epsilon}|\Gamma(\frac{1}{2}+\frac{1}{2}i\epsilon)D_{\nu}(0)|^2 = 
\frac{1}{16\pi\sqrt{w}}e^{-\frac{1}{4}\pi\epsilon}|\Gamma(\frac{1}{4}+\frac{1}{4}i\epsilon)|^2
 \label{a44}
\end{eqnarray}
where a standard expression for $D_{\nu}(0)$ has been used. Of course, one has to be careful in applying the results of \cite{Misner} to the 
present problem. Recall that the parabolic WKB approximation relies on the fact that given a potential which can be adequately represented by a 
parabola near its maxima, there exists classical turning points ($V=E$), and that there is an overlap region, where the parabolic solution and the standard
WKB one are supposed to hold simultaneously. In such a situation, the asymptotic solution of eq.(\ref{a39}) is matched with a WKB solution. In the
present case, this poses a problem. Although classical turning points exist, from eq.(\ref{a34}) we see that 
asymptotically near the horizon, $u_L$ does not have the form of a plane wave, i.e is not of 
a WKB form, whereas the parabolic cylindrical functions asymptote to plane wave solutions in this region. As alluded to before, this is a peculiarity in dealing 
with a naked singularity, which we have to live with. In the region $r_* \to \infty$,  however, such a problem does not arise, and $u_L$ can be obtained 
by matching the parabolic solution with a plane wave \cite{Misner}. We will proceed, with this in mind, i.e our $u_L$ is obtained only from the properties
of the solution at spatial infinity, and we are unable to comment on its behavior near $r=b$. \footnote{An alternative method would be to obtain
a numerical solution for $u_L$ with the boundary conditions specified in eq.(\ref{a34}), as was done for the AdS Schwarzschild black hole in 
\cite{cardoso}. Unfortunately, however, we found that in our case, such numerical solutions
generated with standard Mathematica or Python numerical routines could not be trusted beyond moderately large values of $l$ and $m$.}

Since we now have the expressions for the wave function, using eq.(\ref{a44}) and eq.(\ref{a28}) as inputs in eq.(\ref{a38}), we can calculate the power 
radiated. Of course, we also need the expressions for $\omega_0$ and $u^0$, given by 
eqs.(\ref{a11}) and (\ref{a21}), respectively. In our case, we take the Schrodinger-like equation to be of the form of eq.(\ref{a39}), with $E=0$. 
That is, the potential is taken (from eq.(\ref{a27})) as 
\begin{equation}
V(r) = \left(1-\frac{b}{r}\right)\left(\frac{b}{r^3}+\frac{l(l+1)}{r^2}\right)-m^2\omega_0 ^2\left(1-\frac{b}{r}\right)^{2(1-\nu)}
\label{mdeppot}
\end{equation}
with $\omega_0$ calculated from eq.(\ref{a11}).
The  analogy with quantum mechanical scattering theory is not clear, but we are able to solve the equation nonetheless. 
We note here that there is an immediate caveat in using the potential of eq.(\ref{mdeppot}) in a WKB scheme, namely, the peak of the potential, 
and the value of $\omega_0$ is dependent on the azimuthal quantum number, $m$. 
We thus have to resort to a reasonable approximation, so that the parabolic WKB method can be trusted. 
This can be done only if the peak of the potential does not shift appreciably, on varying $m$ and $l$. We will justify this in sequel. 

Before we start our analysis on the power spectrum, let us first state what we expect. From a gravitational lensing perspective, it can be shown
(see, e.g. \cite{Perlick} and references therein) that for $\nu > {1 \over 2}$, the JNW naked singularity presents qualitatively same features as a black hole. Physically, 
we expect our analysis of the power to retain the same features, i.e for this range of $\nu$, the radiation should be synchrotron in nature, 
(i.e, peak at large $m$), and moreover,
the radiated power should sharply peak in a narrow band of frequencies - for black holes, it is well known \cite{Misner} that at the peak of the spectrum,
more than $99$ percent of the power is radiated in the $l=m$ mode. However, as we will see, as far as the radiated power is concerned, there are interesting 
deviations from this picture, and that only for $\nu > {2 \over 3}$ does the power spectrum have the same qualitative features as a Schwarzschild background.  
A word about the nature of orbits is in order. We will, in sequel, consider both stable and unstable circular orbits in the JNW geometry. The work of \cite{Misner}
specifically focuses on unstable i.e highly relativistic orbits in the Schwarzschild geometry, where the radiation was shown to be synchrotron. 
That such orbits might not be of relevance in astrophysical scenarios was pointed out in \cite{davis}. In what follows, we will not be concerned with this latter fact, 
and our purpose would be to distinguish between black hole and naked singularity backgrounds, as far as radiation properties are concerned. 
\footnote{In the treatment of \cite{Misner}, it is not possible to analytically deal with stable orbits in the Schwarzschild background, since the analogue of the potential
of eq.(\ref{mdeppot}) in that case peaks close to $r=3M$, the location of the photon sphere, whereas stable orbits in the Schwarzschild geometry 
occur beyond $r=6M$, $M$ being the Schwarzschild mass. In our case, it is possible to understand radiation from stable orbits. Expectedly, as we will see, 
this radiation has a large angular spread.} 

\begin{figure}[t!]
\begin{minipage}[b]{0.5\linewidth}
\centering
\includegraphics[width=2.8in,height=2.3in]{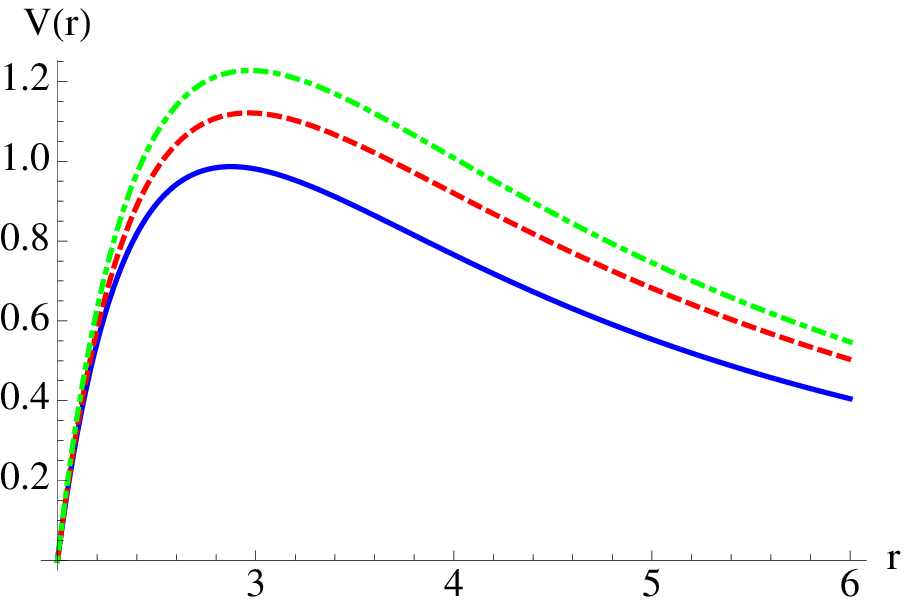}
\caption{$V(r)$ of eq.(\ref{mdeppot}) vs $r$ with $\nu = 0.03$. The solid blue, dashed red and dot-dashed green lines are for 
$m=1, l=1$, $m=5, l=5$ and $m=100,l=100$ respectively.}
\label{v03}
\end{minipage}
\hspace{0.2cm}
\begin{minipage}[b]{0.5\linewidth}
\centering
\includegraphics[width=2.8in,height=2.3in]{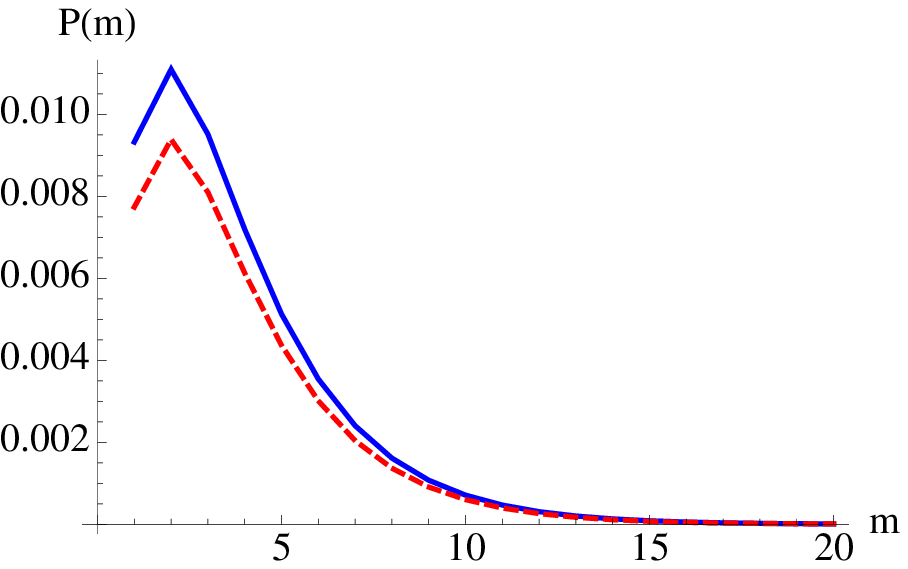}
\caption{Power radiated in the $m$th mode with $\nu = 0.03$. The dashed red line is for the $l=m$ mode, while the solid blue
line is the power, where $l$ has been summed from $m$ to $m+2$.
 }
\label{p03}
\end{minipage}
\end{figure}
To begin with, we consider the case $0<\nu<{1 \over \sqrt{5}}$. We first choose $\nu = 0.03$, so that we are far from the Schwarzschild limit. Recall that 
for this value of $\nu$, all circular orbits are stable, from eq.(\ref{a18}). We will henceforth choose $b = 2$, so that the corresponding Schwarzschild 
mass is set to unity, in appropriate units. We will also set to unity the mass $m_0$ and the scalar charge $q_s$ of eq.(\ref{a22}). 
First, in eq.(\ref{mdeppot}), we substitute $\omega_0$ from eq.(\ref{a11}). Then, in order for the parabolic WKB approximation
to remain valid, we require the particle orbit to be close to the peak of the potential, i.e where $\frac{dV}{dr}|_{r=r_0} = 0$. 
While this condition is satisfied for two 
positive values of $r_0$, one of these is always at a position $r_0 < 2$, and hence discarded. In the present case, we find that
for $(m,l)$ = $(1,1)$, $V(r)$ peaks at $r_0 = 2.88$, while for large values of $m=l$, the peak asymptotes to $r_0 \sim 2.97$. The shift in the peak with 
increase of $m,~l$ is thus small, and we choose our circular orbit at $r_0 = 2.92$, and apply the parabolic WKB approximation. In fig.(\ref{v03}), we have
shown the potential of eq.(\ref{mdeppot}), for $\nu = 0.03$. In this figure, the solid blue, dashed red and dot-dashed green lines correspond to 
$(m,l)=$ $(1,1)$, $(5,5)$ and $(100,100)$ respectively, where we have appropriately scaled $V(r)$, so that the three cases can be displayed on a
single graph. It can be seen that the peak of the potential shifts towards $r \sim 2.97$, as we increase the values of $m=l$. 
In fig.(\ref{p03}), we show our result for the power radiated, as a function of the azimuthal quantum number $m$. The dashed red line 
correspond to the power radiated in the $m=l$ mode, whereas in the solid blue line, we have summed over $l$ from $m$ to $m+2$. It can be seen that
the radiation is dominated by the small $m$ modes. 

Here and later in the paper, we focus on the modes close to $l=m$. This is justified by taking a concrete example. 
In the present case, i.e for $\nu = 0.03$, given a fixed value of $m$, 
the peak of the potential of eq.(\ref{mdeppot}) asymptotes to $\sim 2.99$ when $l$ is made very large, so that our approximation of $r_0 = 2.92$ can still
be applied with reasonable confidence. Specifically, for $l=m=2$, the potential peaks at $r_0 = 2.93$, while for $(l,m)$ = $(2,3)$ and $(2,4)$, the potential
peaks at $r_0 = 2.96$ and $2.98$ respectively. Now, for $l=m=2$, the numerical value of the radiated power is $P_{l=m=2} \sim 9.4 \times 10^{-3}$, while
$P_{l=4,m=2} \sim 1.7 \times 10^{-3}$ ($P_{l=4,m=3} = 0$ from properties of spherical harmonics). Thus, the radiated power decreases away from
the $l=m$ mode, and to a good approximation, it is enough to consider modes near $l=m$, to understand the nature of the spectra. 
We will keep this in  mind in our future discussions. 

\begin{figure}[t!]
\begin{minipage}[b]{0.5\linewidth}
\centering
\includegraphics[width=3.0in,height=2.3in]{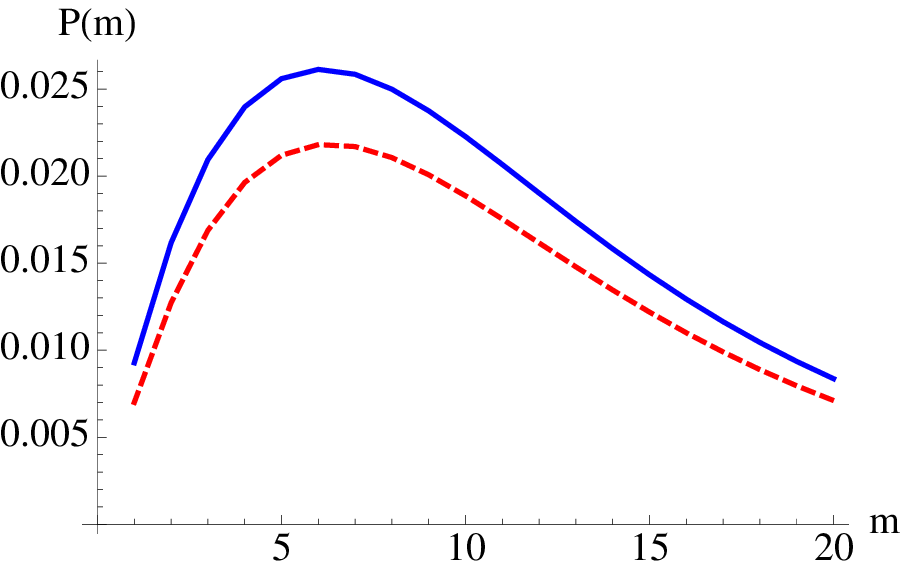}
\caption{Power radiated as a function of $m$ for $\nu = 0.6$, with $r_0 = 0.6$.The solid blue curve is the case where $l$ has been summed from $m$ to
$m+2$. The dashed red curve is the $l=m$ mode. }
\label{06small}
\end{minipage}
\hspace{0.2cm}
\begin{minipage}[b]{0.5\linewidth}
\centering
\includegraphics[width=3.0in,height=2.3in]{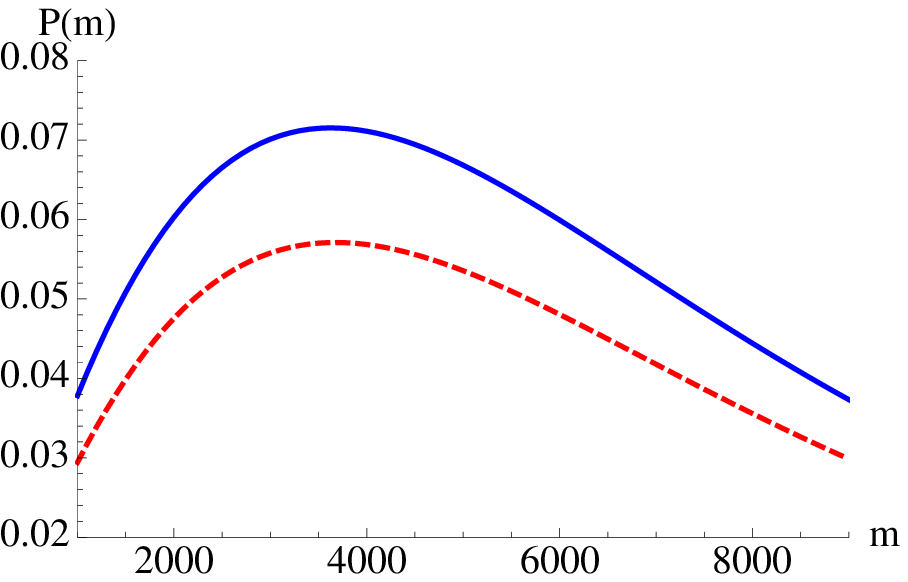}
\caption{Power radiated as a function of $m$ for $\nu = 0.6$, with $r_0 = 2.201$. For the solid blue curve, $l$ has been summed from $m$ to $m+2$. The
dashed red curve is the $l=m$ mode.}
\label{06large}
\end{minipage}
\end{figure}
We now come to the region ${1 \over \sqrt{5}} < \nu<0.5$. Here, we will take $\nu = 0.47$ as our representative value. 
We choose $r_0 = 2.7$ in our analysis below, and it can be checked that the parabolic WKB approximation can be well justified for this choice, for
small values of $m$ and $l$. This will be enough for us, as the radiated power is negligible for large values of $m$ and $l$.  
Note that from eq.(\ref{a17}),  this means that our particle is in the unstable region, as stable orbits exist for $r_0 < 2.087$ and $r_0 > 2.733$, and 
we have kept the particle very close to the boundary of the unstable region. Here, we find that for the $l=m$ modes, the radiated power 
maximizes at small values ($\sim 4$) of $m$. The main features of the radiation spectrum is similar to the previous case, $0<\nu<{1 \over \sqrt{5}}$,
i.e it is dominated by the small $m$ modes, and we will not elaborate on this further. \footnote{In hindsight, we note here that when $\nu = 0.5^-$, an 
interesting possibility occurs, the radiation becomes synchrotron, and can resemble those with black hole backgrounds. This will be clear from our
discussion below. }

While the cases considered till now correspond to strongly naked singularities, more interesting cases occur for $0.5 < \nu < 1$, for which a 
photon sphere exists. As an illustration, let us take $\nu = 0.6$. Here, the radius of the photon sphere is 
$r_{\rm ps} = 2.2$, and from eq.(\ref{a17}), $r_+ = 3.69$ and $r_- = 1.91$. Thus, stable circular orbits will exist only for $r_0 > 3.69$. In this case, 
as we show below, there are two distinct peaks of the potential, both in the unstable region, one of which is very close
to the photon sphere. Specifically, for $\nu = 0.6$, the potential of eq.(\ref{mdeppot}) extremizes, for $m=l=2$, at $r_0 = 2.70$ and at $r_0 = 1.8$, the latter
being discarded (we remind the reader that we are interested in the modes near $m=l$). As we increase to $m=400$ (=$l$), there are two maxima,
at $r_0 = 2.41$ and at $r_0 = 2.19$. Further increasing the value of $m$, we find that one of the maxima asymptote to $r_0=2.4$ and the other merges
with the radius of the photon sphere. We can thus consider unstable orbits very close to $r_{\rm ps}$, 

Now, for small $m$ and $l$, we choose our orbit at $r_0 = 2.6$, where we expect our parabolic WKB approximation to hold good. In fig.(\ref{06small}), we
have shown the spectrum as a function of $m$, and see that the peak of the spectrum is at $m \sim 7$. In this figure, the solid blue curve denotes the case for
which $l$ has been summed from $m$ to $m+2$. The dashed red curve is the $l=m$ mode. It can be checked that taking $r_0 \sim 2.4$, i.e close to one of
the asymptotic values of the potential maximum, does not change the result appreciably. Fig.(\ref{06large}) shows the spectrum for $r_0 = 2.201$, where
the same color coding as in fig.(\ref{06small}) has been used. The peak of the spectrum has changed drastically from the previous case. This,
in fact, closely resembles a black hole spectrum, where it is known that unstable orbits exhibit synchrotron radiation \cite{Misner}. 

As we increase the value of $\nu$, we find an additional feature in the power spectrum. Namely, at $\nu = 2/3$, the two peaks of the potential for large
values of $m$, as alluded to above, merge. Specifically, for $\nu=2/3$, for $l=m=2$, the potential of eq.(\ref{mdeppot}) has maxima at $r_0 = 2.71$ and $1.89$, 
with the second one being unphysical. As we increase the value of $m$ ($=l$), the two maxima merge at $r_0 = 7/3$. For $\nu > 2/3$, we find that there is a single
maximum of the potential in the region $r > 2$, which asymptotes to $r_{\rm ps}$ as the values of $m$ and $l$ are increased.
While the radiated power maximizes for small to moderate values of $m$ for orbits away from $r_{\rm ps}$ in these cases, 
\footnote{We always stay within the validity of the parabolic WKB approximation.} it shows synchrotron behavior
when the particle radius is close to $r_{\rm ps}$. As expected, in the limiting case $\nu \to 1$, the potential always maximizes close to $r_{\rm ps}$,
for all values of $m$ and $l$. 
\begin{figure}[t!]
\begin{minipage}[b]{0.5\linewidth}
\centering
\includegraphics[width=3.0in,height=2.3in]{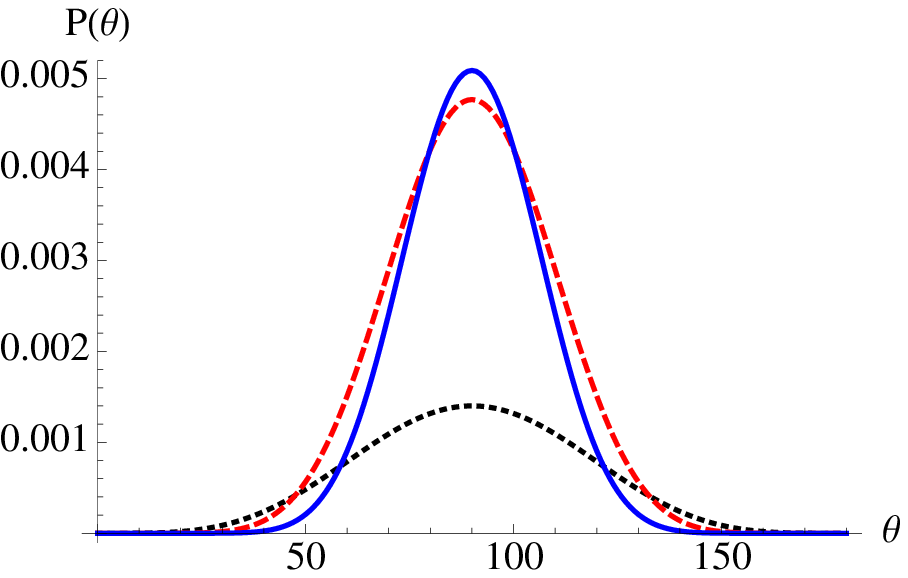}
\caption{Power radiated as a function of $\theta$ in the (small) $m=l$  mode, for $\nu = 0.03$ (dotted black), $0.47$ (dashed red) and $0.6$ (solid blue).}
\label{theta1}
\end{minipage}
\hspace{0.2cm}
\begin{minipage}[b]{0.5\linewidth}
\centering
\includegraphics[width=3.0in,height=2.3in]{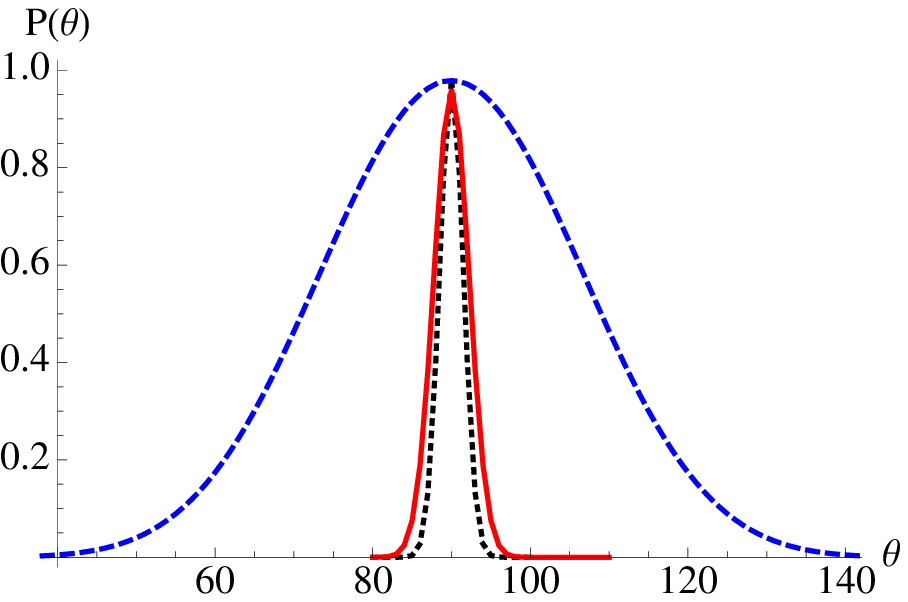}
\caption{Power radiated as a function of $\theta$ for large $m$, for $\nu = 0.6$ (dotted black) and $0.9$ (solid red), and for $\nu = 0.6$, with small $m$ shown in dashed blue.}
\label{theta2}
\end{minipage}
\end{figure}

We also comment upon a further distinct feature of the power spectrum. To illustrate the point, 
we take $\nu = 0.6$, and choose $r_0 = 2.201$, close to the radius of the photon sphere. Here, the peak of the spectrum occurs at $m = 3669$. We find that 
the ratio of the $l=m$ and the $l=m + 2$ modes is $\sim 4$. In contrast, for $\nu = 0.9$, choosing $r_0 = 2.801$ ($r_{\rm ps} = 2.8$) the peak power is
for $m = 1448$, and the same ratio yields a value $\sim 24$. Hence, as we approach the black hole limit $\nu \to 1$, most of the power is radiated in the 
$l=m$ mode, consistent with the results of \cite{Misner}, but away  from it, although the radiation might be synchrotron in nature, a few modes away from $l=m$ 
cannot be entirely neglected. 

Finally, we have also studied the angular dependence of the power spectrum, at its peak, for various values of $\nu$. The results are summarized in figs.(\ref{theta1}) and
(\ref{theta2}). In fig.(\ref{theta1}), we consider the $m$ mode for which the power is maximized, for small $m$, and $l=m$.
We see that the radiation is spread over a wide angular region, for  $\nu = 0.03$ (dotted black), $0.47$ (dashed red) and $0.6$ (solid blue). 
In contrast, the dotted black and solid red lines in fig.(\ref{theta2}) show the angular dependence of the maximally radiating mode for large $m$ (i.e 
when the particle is close to the photon sphere), for $\nu = 0.6$ and $0.9$ respectively. We see a relatively small angular spread of the spectrum. 
To make the contrast clearer, we have plotted, in fig.(\ref{theta2}), the small $m$ case of $\nu = 0.6$, shown in dotted blue. 

\section{Discussion and Summary}
 
In this paper, using a WKB approximation scheme following \cite{Misner}, we have made a comprehensive investigation of scalar radiation in 
the background of the JNW naked singularity. Our main conclusion here is that the nature
of the power spectrum reveals a rich structure, for different ranges of the parameter $\nu$ appearing in the JNW metric. For $\nu < {1 \over \sqrt{5}}$, the 
spectrum shows a peak at small values of the azimuthal quantum number. The situation is unchanged for ${1 \over \sqrt{5}} < \nu < {1 \over 2}$. However, 
from our discussion, it follows that in this region, close to $\nu = 0.5^-$, the nature of the spectrum may become synchrotron. The most interesting
case occurs for ${1 \over 2} < \nu < 1$, where we saw that upto $\nu = {2 \over 3}$, there are two maxima of the potential of eq.(\ref{mdeppot}). 
Thus, there might be two distinct types of radiation spectra, one of them closely resembling
that from a black hole, when the particle is close to the photon sphere. Beyond this value of $\nu$, the potential has a single maximum, which stays
close to the photon sphere as $\nu \to 1$. 
We have also seen that away from the $\nu \to 1$ limit, when the radius of the particle orbit is close to that of the photon sphere, although
the qualitative behavior of the spectrum may be similar to a black hole background, there is an
important difference. Namely that a few $l$ modes away from $l=m$, contributes significantly to the power, in contrast to the black hole case, where
more than $99$ percent of the power is radiated in the $l=m$ mode. 

The nature of naked singularities is far from being fully understood, and we believe that our results complement existing ones in the literature, 
that study the similarities and differences between these objects and black holes. It will be interesting to study the nature of electromagnetic or 
gravitational radiation spectra in the background of the JNW singularity. We leave this for a future publication.

\begin{center}
{\bf Acknowledgements}
\end{center}
It is a pleasure to thank V. Cardoso and V. Suneeta for useful email correspondence.

\end{document}